\documentclass[journal=nalefd, layout=twocolumn]{achemso}

\usepackage[T1]{fontenc} 

\usepackage{amsmath,color, tikz}
\usetikzlibrary{matrix}
\usepackage{amssymb}
\usepackage{amsfonts}
\usepackage{amsthm}
\usepackage{mathtools}
\usepackage{comment}
\usepackage{hyperref}

\hypersetup{
	colorlinks=true,
	citecolor=blue,
	linkcolor=blue,
	urlcolor = blue
}

\usepackage{xspace}
\usepackage{cleveref}

\usepackage{algorithm}
\usepackage{algpseudocode}
\usepackage{dsfont}

\usepackage{bbold}
\usepackage{chemformula}
\usepackage{siunitx}

\DeclarePairedDelimiterX\braket[2]{\langle}{\rangle}{#1 \delimsize\vert #2}
\DeclarePairedDelimiterX\braket3[3]{\langle}{\rangle}{#1 \delimsize\vert #2 \delimsize\vert #3}

\usepackage{accents}

\usepackage{fancyvrb}

\newcommand{\dbtilde}[1]{\accentset{\approx}{#1}}

\newcommand{\vmu}{\boldsymbol{\mu}}

\newcommand{\vR}{\mathbf{r}}

\newcommand{\vk}{\mathbf{k}}

\newcommand{\avg}[1]{\left\langle #1\right\rangle}

\newcommand{\MaxwellLink}{\textsc{MaxwellLink}\xspace}

\usepackage{pdfpages}
\makeatletter
\AtBeginDocument{\let\LS@rot\@undefined}
\makeatother

	\title{Nonlinear Freezing of Vibrational Polariton Transport via Mesoscale Simulations}

    \author{Xinwei Ji}
    \affiliation{Department of Physics and Astronomy, University of Delaware, Newark, Delaware 19716, USA}
	
	\author{Tao E. Li}%
	\email{taoeli@udel.edu}
	\affiliation{Department of Physics and Astronomy, University of Delaware, Newark, Delaware 19716, USA}
    
\begin{document}

    \begin{tocentry}
    \centering
        \includegraphics[width=0.6\linewidth]{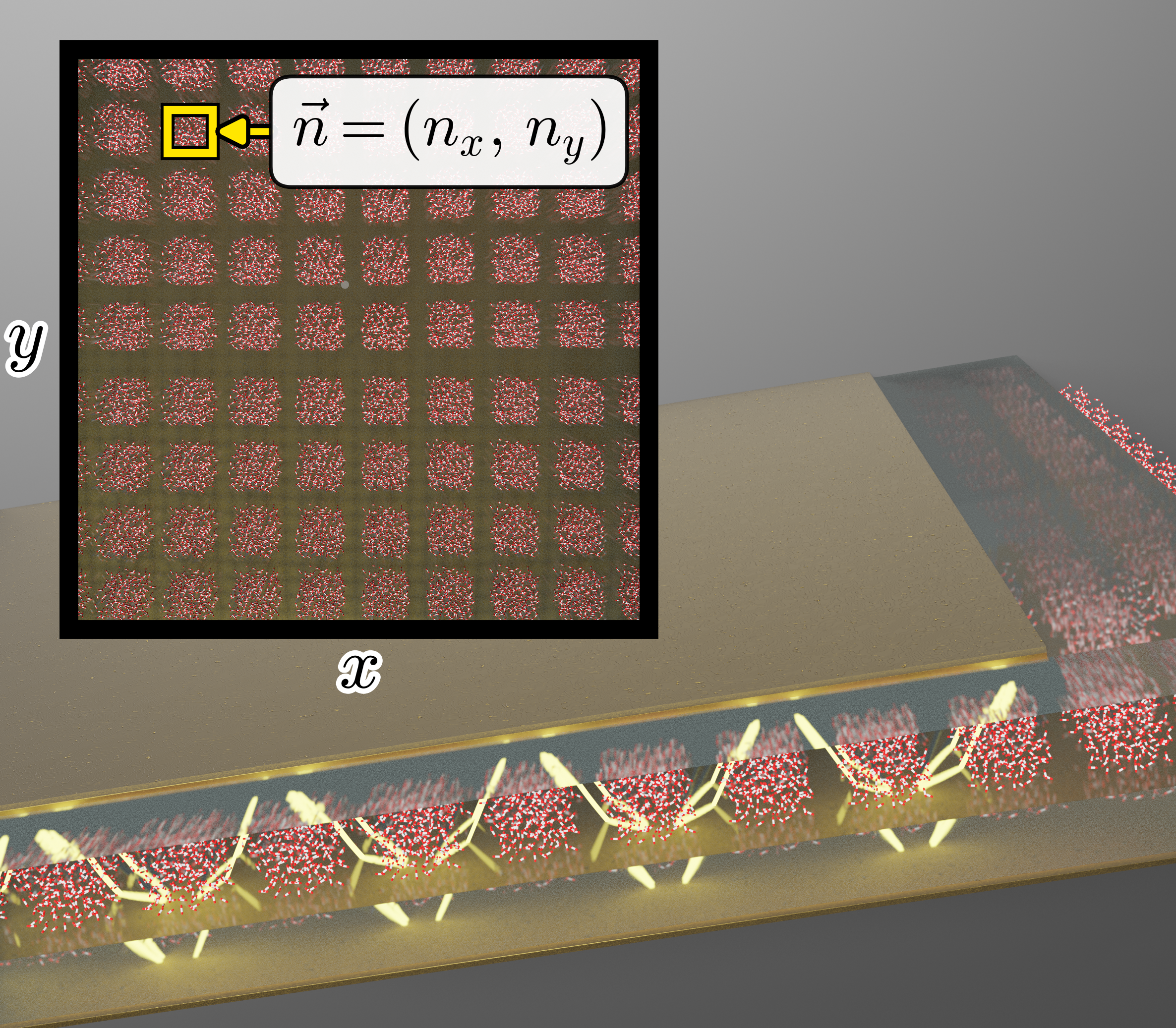}
    \end{tocentry}

    \begin{abstract}
    Two-dimensional real-space imaging of vibrational polariton transport in planar Fabry--P\'erot microcavities is numerically simulated via the mesoscale cavity molecular dynamics approach, which self-consistently propagates $\sim\!2\times10^4$ realistic molecular simulation cells on a two-dimensional grid coupled to the same number of cavity modes. Beyond the well-known polariton ballistic-to-diffusive turnover in the linear response regime, these atomistic simulations reveal a nonlinear freezing mechanism of vibrational polariton transport, i.e., under strong pumping of the upper polariton, the initially ballistically propagating upper polariton completely freezes and localizes energy to molecules at specific locations. This mechanism originates from pump-induced breaking of the in-plane translation symmetry:  significant molecular excitations at the pulse hot spot broaden the polariton density of states, thus funneling population to the $k_{\parallel}\rightarrow 0$ band edge with vanishing group velocities. Overall, our work provides numerical insights for exploring novel mechanisms of vibrational polaritons.
    \paragraph{Keywords:} Vibrational Strong Coupling, Polariton Transport, Molecular Dynamics, Strong Pumping.
	\end{abstract}

	\maketitle

    Vibrational polaritons form when a vibrational mode of a large ensemble of molecules couples strongly to an optical or plasmonic microcavity \cite{Shalabney2015,Long2015,Wright2023,Brawley2025NatChem}. In this vibrational strong coupling (VSC) regime, experiments report modified thermally-activated chemical reaction rates and crystallization processes \cite{Thomas2016, Thomas2019_science,Hirai2020Crys,Ahn2023Science}, enhanced remote energy transfer \cite{Xiang2020Science}, and altered infrared (IR) photochemistry \cite{Chen2022,Yin2025}. These experiments raise an intriguing question: How can vibrational polaritons significantly impact local molecular processes, with or without external pumping, even though the light-matter coupling per molecule is intrinsically small for most cavity setups \cite{Ribeiro2018,Li2022Review,Simpkins2023,Mandal2023ChemRev,Ruggenthaler2023,Xiang2024}? Addressing this question poses significant challenges to existing theoretical models and numerical toolboxes, as both mode-resolved cavity structures and realistic molecular disorder must be properly treated \cite{Fregoni2022,Poddar2026}, not to mention the necessity of approaching the large $N$ limit to distinguish collective molecular response from predictions under single- or few-molecule strong coupling \cite{Perez-Sanchez2023,Kotov2025}.

    Recent two-dimensional (2D) real-space imaging experiments of exciton-polariton \cite{Xu2023Polariton,Balasubrahmaniyam2023} and vibrational polariton transport \cite{Bhakta2026} reveal ballistic-to-diffusive transport along the lateral direction of planar Fabry--P\'erot cavities: Upon spatially selective pumping at a well-defined in-plane momentum ($k_{\parallel}$), the excited upper polariton (UP) initially propagates ballistically  with a fixed group velocity, followed by diffusive transport arising from molecular disorder and dephasing \cite{Xu2023Polariton,Balasubrahmaniyam2023,Bhakta2026}. This ballistic-to-diffusive turnover has been studied using both quantum-mechanical calculations and  analytical treatments \cite{Sokolovskii2022tmp,Ribeiro2022,Suyabatmaz2023,Zhou2023,Engelhardt2023,Blackham2025,Fowler-Wright2025,Chng2026,Liu2025}. Despite these exciting advances, atomistically resolved 2D real-space imaging of vibrational polariton dynamics remains challenging. This gap hinders direct microscopic understanding of how propagating polaritons localize their energy to individual molecules, a key question for resolving the mysteries in polariton chemistry experiments. 

\begin{figure*}
		\centering
		\includegraphics[width=0.7\linewidth]{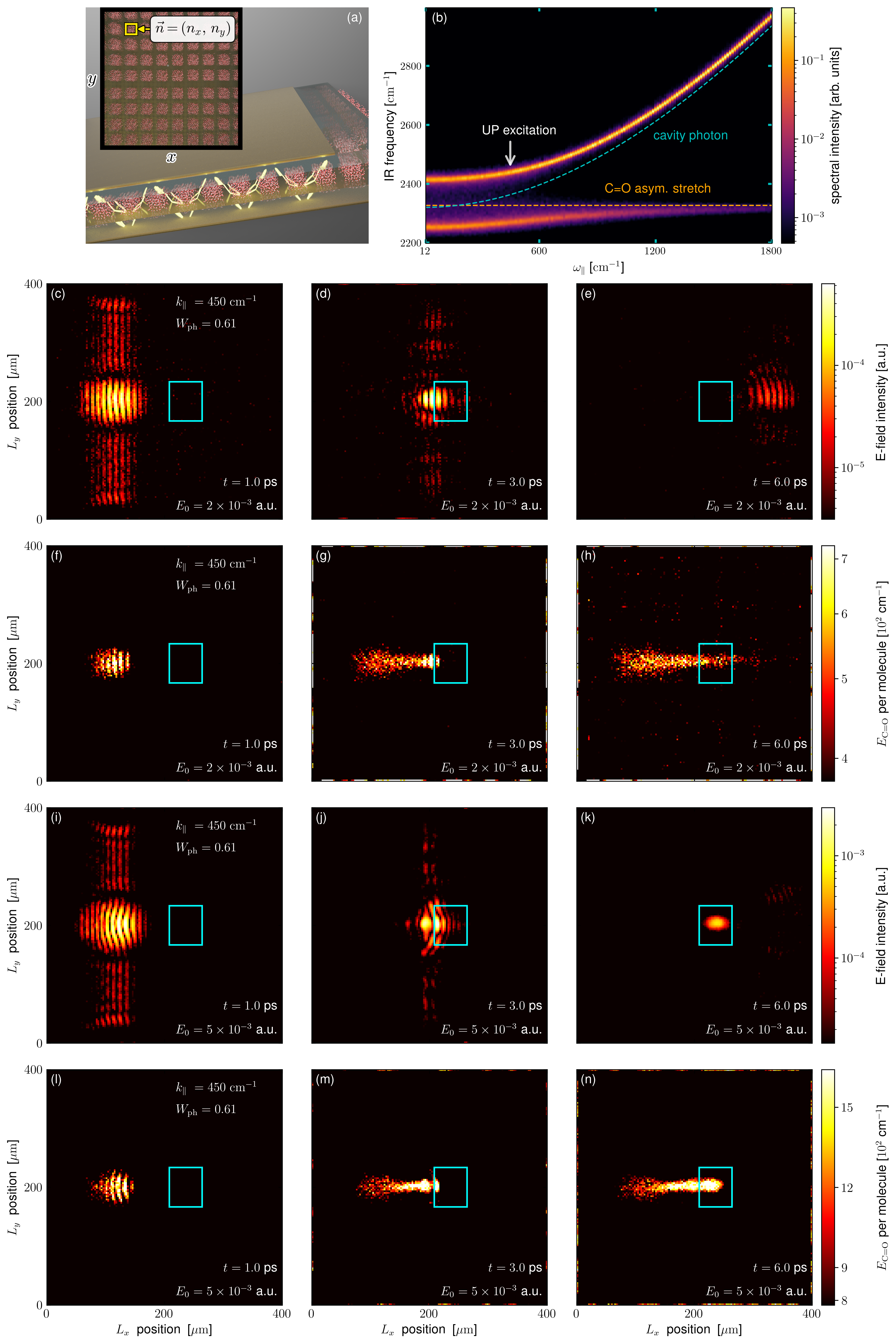}
		\caption{Simulated 2D real-space imaging of vibrational polariton transport. (a) A cavity setup consisting of a grid of $144\times144$ points, each representing an independent LAMMPS MD engine, is coupled to the same number of cavity modes along the $xy$ plane. (b) Simulated linear dispersion relation. (c-h)  Snapshots of the (c-e) E-field intensity distribution and (f-h) per-molecule \ch{C=O} energy distribution during UP transport under a weak pulse excitation (amplitude $E_0=2\times 10^{-3}$ a.u.) at $t=1.0, 3.0$, and 6.0 ps, respectively. (i-n) Analogous E-field intensity distribution and per-molecule \ch{C=O} energy distribution under a strong pulse excitation (amplitude $E_0 = 5\times 10^{-3}$ a.u.). Cyan squares indicate the polariton freezing location under strong pumping. Here, $144\times 144$ molecular grid points are coupled self-consistently with the same number of photon modes, resulting in approximately 2.2 million explicitly simulated atoms. See the SI for simulation details.
		}
		\label{fig:2d_imag}
    \end{figure*}

    Here, we report atomistic simulations of 2D real-space imaging of vibrational polariton dynamics in planar Fabry--P\'erot cavities. With these simulations, we not only characterize the ballistic-to-diffusive turnover in the linear response regime, by also reveal a novel mechanism of UP transport under strong pumping: from ballistic propagation to \textit{nonlinear freezing}. Our calculations employ  the recently developed mesoscale cavity molecular dynamics (CavMD) approach \cite{Li2024CavMD}, which involves the self-consistent Maxwell-MD simulation of up to $20,736$ molecular simulation cells (each an independent LAMMPS MD engine \cite{Thompson2022}) coupled to the same number of cavity normal modes on a 2D spatial grid; see Fig. \ref{fig:2d_imag}a for the simulation setup and Supplemental Information (SI) for simulation details.

    In brief, this mesoscale CavMD approach simulates the photonic dynamics via classical 2D Maxwell's equations in the normal mode basis \cite{Li2024CavMD}:
    \begin{equation}
        m_{k\lambda} \ddot{\dbtilde{q}}_{k\lambda} = - m_{k\lambda} \omega_{k}^2\dbtilde{q}_{k\lambda} - \widetilde{\varepsilon}_{k\lambda} d_{\text{g},k\lambda} , \label{eq:EOM_new_ph}
    \end{equation}
    where $m_{k\lambda}$, $\omega_{k}$, $\dbtilde{q}_{k\lambda}$, and $\dbtilde{p}_{k\lambda}$ denote the auxiliary mass, frequency, position, and momentum for the cavity photon mode polarized along direction $\lambda=x,y$ at wave vector $k=|\vk|$. The quantity $\widetilde{\varepsilon}_{k\lambda}$ characterizes the effective light-matter coupling strength between each cavity mode and molecule, and $d_{\text{g},k\lambda}=\sum_{\vec{n}} \vmu_g^{\vec{n}}  \cdot \sqrt{\mathcal{V}}\mathbf{f}_{k\lambda}(\vR_{\vec{n}})$ represents the molecular dipole moment coupled to the cavity photons. Here, $\vec{n}=(n_x,n_y)$ spans the 2D spatial grid (Fig. \ref{fig:2d_imag}a inset); $\vmu_g^{\vec{n}}$ is the molecular dipole vector of the MD cell at grid point $\vec{n}$; $\mathcal{V}$ denotes the effective cavity volume; and $\mathbf{f}_{k\lambda}(\vR_{\vec{n}})$ is the photonic mode function at $\vR_{\vec{n}}$, the spatial coordinate of each molecular grid point. For a planar Fabry--P\'erot cavity, we adopt the mode functions  corresponding to the midplane of a rectangular parallelepiped.

    At each grid point $\vec{n}$, mesoscale CavMD evolves classical Newtonian equations of motion for all atoms in the MD simulation cell \cite{Li2024CavMD}:
    \begin{equation}
        M_{\alpha} \ddot{\mathbf{R}}_{\alpha} = \mathbf{F}_{\alpha}^{(0)} + \mathbf{F}_{\alpha}^{\rm cav} ,
    \end{equation}
    where $M_{\alpha}$, $\mathbf{R}_{\alpha}$, $\mathbf{F}_{\alpha}^{(0)}$, and $\mathbf{F}_{\alpha}^{\rm cav}$ denote the mass, position, bare nuclear force, and cavity force on each nucleus $\alpha$. In the cavity force  expression 
    $\mathbf{F}_{\alpha}^{\text{cav}} = -\sum_{k\lambda} \left(
                \widetilde{\varepsilon}_{k\lambda} \dbtilde{q}_{k\lambda} + 
                \frac{ \widetilde{\varepsilon}_{k\lambda}^2}{m_{k\lambda} \omega_{k}^2} d_{\text{g},k\lambda}
                \right) \frac{\partial d_{\text{g},k\lambda} }{\partial \mathbf{R}_{\alpha}}$, the term proportional to $\widetilde{\varepsilon}_{k\lambda}^2$ stems from the dipole self-energy term in the light-matter Hamiltonian. In other words, the  dipole self-energy term\cite{flickAtomsMoleculesCavities2017b,flickKohnShamApproach2015} is incorporated in CavMD dynamics.
    We implement this mesoscale CavMD approach in the latest \MaxwellLink package \cite{Ji2026MaxwellLink}, which enables tens of thousands of LAMMPS molecular engines to be concurrently coupled to the multimode cavity at affordable computational cost.

    Fig.  \ref{fig:2d_imag}b shows the equilibrium polariton dispersion relation for liquid \ch{CO2}  confined in a planar Fabry--P\'erot cavity at 300 K, computed with mesoscale CavMD.   The setup comprises $144\times 144$ molecular grid points evenly placed along the $xy$ plane of a $400\times 400$ $\mu$m$^2$ area (Fig.  \ref{fig:2d_imag}a). At each grid point, 36 liquid-phase \ch{CO2} molecules are explicitly simulated under periodic boundary conditions using non-polarizable, anharmonic force fields \cite{Cygan2012,Li2020Nonlinear}, amounting to the explicit simulation of approximately 2.2 million atoms. The effective light-matter coupling strength per molecule is $\widetilde{\varepsilon}=2.083\times10^{-6}$ a.u. The cavity modes have frequencies at  $\omega_{k} = \sqrt{\omega_{\perp}^2 + (l_x\pi c/L_x)^2 + (l_y\pi c/L_y)^2}$, where $\omega_{\perp}=2320$ cm$^{-1}$ is at resonance with the \ch{C=O} asymmetric stretch mode, the integers $l_{x,y} \in [1, 144]$, and $L_x=L_y=400$  $\mu$m. A cavity lifetime of 5 ps is applied by attaching the cavity modes to a Langevin thermostat, while molecules are not directly attached to any thermostat. The E-field propagation is absorbed at the geometry boundary to avoid reflection \cite{Li2018Spontaneous}. Note that liquid \ch{CO2} molecular system under VSC has been numerically studied previously with CavMD \cite{Li2021Relaxation,Li2020Nonlinear,Li2021Collective,Li2021Solute}, which shows qualitative agreement with  ultrafast vibrational polariton experiments \cite{Xiang2018,Xiang2019}. 

    We then  excite the UP at $k_{\parallel}=450$ cm$^{-1}$ along the $x$-direction using a Gaussian pulse with temporal width $\tau=0.5$ ps and spatial width $\Delta x = 120$ $\mu$m. This UP possesses a photonic weight of $W_{\rm{ph}}=0.61$ calculated from Hopfield coefficients \cite{Hopfield1958}. At a weak pulse amplitude $E_0=2\times 10^{-3}$ a.u., Figs. \ref{fig:2d_imag}c-e display three snapshots of the E-field intensity over the 2D cavity mirror plane at $t=1.0$, 3.0, and 6.0 ps, respectively. Here, each pixel represents cavity modes coupled to a realistic MD simulation cell.  In Figs. \ref{fig:2d_imag}c and d, the propagating UP gradually loses the Gaussian envelope and dephases, because the molecular component of  the polariton interacts directly with local molecules (or molecular dark states) via intermolecular interactions. Correspondingly, as demonstrated in Fig. \ref{fig:2d_imag}f-h, the propagating UP continuously deposits energy to local molecular vibrations, leading to a real-space molecular excitation pattern which resembles pump-probe microscopy experiments of exciton-polaritons \cite{Balasubrahmaniyam2023}.

    Under a stronger Gaussian pulse ($E_0=5\times10^{-3}$ a.u.)\cite{Li2020Nonlinear}, as shown in Figs. \ref{fig:2d_imag}i and j, the early-time E-field dynamics ($t\leq 3$ ps) remain similar to those at the weak-excitation limit (Figs. \ref{fig:2d_imag}c and d). By contrast, the later-time behavior differs markedly: at $t=6$ ps (Fig. \ref{fig:2d_imag}k), the E-field remains localized around the centroid position at $t=3$ ps (cyan square), whereas the UP continues propagating under weak excitation (Fig. \ref{fig:2d_imag}e). As a result, under strong pumping, local molecules at this E-field localization position become significantly excited (Fig. \ref{fig:2d_imag}n). See also supplementary movies for the time-resolved dynamics. Overall, Fig. \ref{fig:2d_imag} not only provides atomistically resolved 2D real-space imaging of vibrational polariton transport, but also reveals the possibility of nonlinear freezing of vibrational polariton transport under strong pumping.

    \begin{figure}
		\centering
		\includegraphics[width=1.0\linewidth]{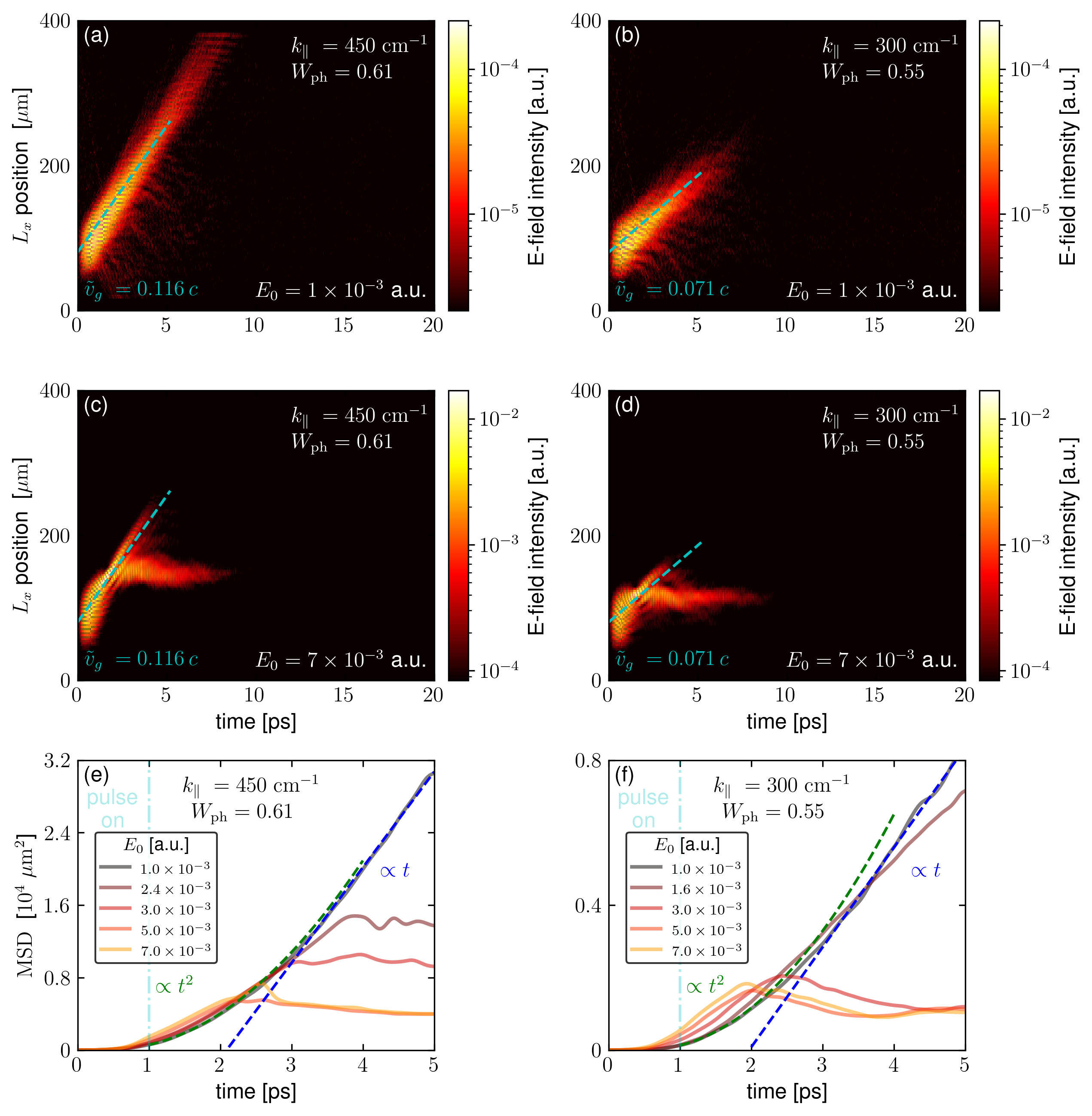}
		\caption{
        UP transport dynamics in a 1D Fabry--P\'erot cavity. (a,b) E-field intensity dynamics when the UP at (a) $k_{\parallel}=450$ or (b) 300 cm$^{-1}$ is weakly excited (amplitude $E_0=1\times 10^{-3}$ a.u.). (c,d) Analogous E-field intensity dynamics under strong pumping (amplitude $E_0=7\times 10^{-3}$ a.u.).
        (e,f) Corresponding mean squared deviation (MSD) of the E-field intensity distribution under increased pulse amplitude $E_0$ (gray to orange). The pumping strength thresholds for nonlinear freezing are approximately (e) $E_0=2.4\times10^{-3}$ a.u. and (f) $E_0=1.6\times10^{-3}$ a.u.
		}
		\label{fig:1d_imag}
    \end{figure}

    To analyze the time-resolved electromagnetic (EM) and molecular dynamics in greater detail, we perform reduced 1D simulations in which the $y$-direction is taken to be homogeneous, aiming to model VSC in a 1D Fabry--P\'erot cavity. In the reduced 1D simulations, 144 molecular grid points are coupled to 144 cavity modes along the $x$-direction, with the simulation parameters following the 2D case.
    
    Figs. \ref{fig:1d_imag}a and c show the time-resolved E-field intensity  after exciting the UP at $k_{\parallel}=450$ cm$^{-1}$ under weak and strong pumping, respectively. At the weak excitation limit (Fig. \ref{fig:1d_imag}a), the UP propagates ballistically  at the analytical group velocity ($\tilde{v}_{\rm{g}}=\partial \omega_{\rm{UP}}/\partial k_{\parallel}$, dashed cyan line; see also SI Fig. S1) \cite{Deng2010}, while small components of the EM field scatter to other directions and the polariton wavepacket gradually dephases. Under strong pumping  (Fig. \ref{fig:1d_imag}c), by contrast, a significant component of the wavepacket ceases to propagate and freezes starting from $t\gtrsim 2$ ps, consistent with the 2D results. 
    
    Fig. \ref{fig:1d_imag}e further plots the mean squared deviation (MSD) of the  E-field intensity dynamics for the $k_{\parallel}=450$ cm$^{-1}$ excitation. At the weak excitation limit (solid gray line), the MSD evolves ballistically (fitted by dashed green line, $\propto t^2$) at initial times, and then transitions to diffusive transport (fitted by dashed blue line, $\propto t$) in later times (t $\gtrsim 3.5$ ps). At larger Gaussian pulse amplitudes ($E_0 \geq 3\times 10^{-3}$ a.u.), clear freezing of the polariton propagation emerges at $t>2$ ps. When the pulse power is enhanced, the nonlinear freezing pathway also emerges slightly earlier.

    The right panel of Fig. \ref{fig:1d_imag} reports the analogous dynamics when the UP at a smaller $k_{\parallel}$ value (300 cm$^{-1}$) is excited. Owing to the smaller photonic weight in the UP ($W_{\rm{ph}}=0.55$), weak-excitation propagation  (Fig. \ref{fig:1d_imag}b) behaves more diffusive than that in Fig. \ref{fig:1d_imag}a and exhibits a ballistic-to-diffusive transition around $t=2.2$ ps (Fig. \ref{fig:1d_imag}f). Capturing this transition agrees with earlier experimental \cite{Bhakta2026} and theory work \cite{Sokolovskii2022tmp,Fowler-Wright2025}, validating mesoscale CavMD for probing vibrational polariton transport.
    Beyond linear response, nonlinear freezing of the UP propagation still occurs under strong pumping (Figs. \ref{fig:1d_imag}d,f).

    SI Fig. S2 additionally shows the analogous 1D E-field dynamics at a few different $k_{\parallel}$ values.  1D simulations with enlarged  molecular systems (by factors of $2^n$, $n=1,2,3$; SI Fig. S3) yield similar transport and freezing behaviors, provided that both the Rabi splitting and energy gain per molecule remains unchanged as those reported in Fig. \ref{fig:1d_imag}c. This $N$-scaling suggests that nonlinear polariton freezing  may persist at the large $N$ limit.

    \begin{figure}
		\centering
		\includegraphics[width=1.0\linewidth]{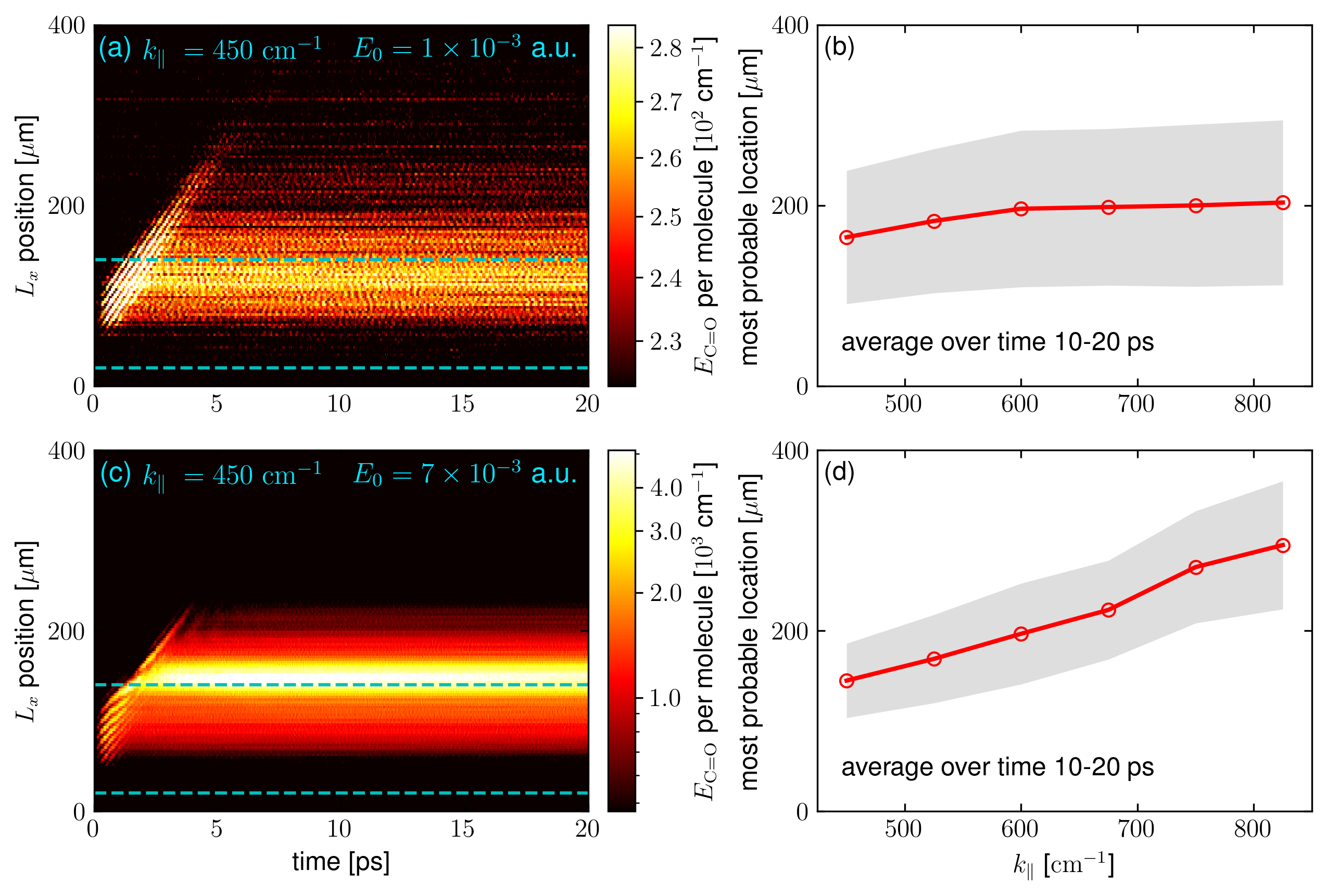}
		\caption{
        (a) Time-resolved \ch{C=O} energy distribution in a 1D Fabry--P\'erot cavity during $k_{\parallel}=450$ cm$^{-1}$ UP transport under weak pumping, corresponding to Fig. \ref{fig:1d_imag}a. The pulse excitation region is marked between dashed cyan lines. (b) Most probable location of the \ch{C=O} energy distribution at later times for different $k_{\parallel}$ excitations. (c,d) Analogous results under strong pumping, corresponding to Fig. \ref{fig:1d_imag}c. 
		}
		\label{fig:remote_energy_transport}
    \end{figure}

    We then focus on the molecular consequence of nonlinear freezing. Figs. \ref{fig:remote_energy_transport}a,c report the corresponding \ch{C=O} bond energy dynamics  after the UP excitation at $k_{\parallel} = 450$ cm$^{-1}$. Under weak excitation (Fig. \ref{fig:remote_energy_transport}a), the propagating UP continuously deposits energy to molecules along the spatial grid due to polariton dephasing into molecular dark modes. Accordingly, the most probable location of the \ch{C=O} bond energy ($E_{\ch{CO}}$) distribution along the spatial grid $\{x_i\}$ remains stationary across different $k_{\parallel}$ excitations (Fig. \ref{fig:remote_energy_transport}b). Here, the most probable location is calculated by 
    $\avg{x}=\sum_i x_i \Delta E_{\ch{CO}}^i / \sum_i \Delta E_{\ch{CO}}^i$, where $\Delta E_{\ch{CO}}^i=E_{\ch{CO}}^i-E_{\ch{CO}}^i(t=0)$ denotes the energy gain per grid point. It is worth noting that, under weak pumping, the molecular energy gain is relatively small, leading to a broad spatial distribution of \ch{C=O} energy (Fig. \ref{fig:remote_energy_transport}b). In other words, the most probable location in this regime does not provide a meaningful measure of polariton localization. 
    
    Under strong pumping, Fig. \ref{fig:remote_energy_transport}c shows strong accumulation of \ch{C=O} bond energy at the polariton freezing location (at $x\approx150$ $\mu$m; cf. Fig. \ref{fig:1d_imag}c). Scanning the UP excitation over different $k_{\parallel}$ values (Fig. \ref{fig:remote_energy_transport}d; see also SI Fig. S4), the most probable location of \ch{C=O} bond energy distribution ($\avg{x}$) is proportional to the in-plane wave number $k_{\parallel}$. This suggests that the spatial location of remote molecular energy deposition can be controlled by tuning $k_{\parallel}$ excitations. 

    \begin{figure}
		\centering
		\includegraphics[width=1.0\linewidth]{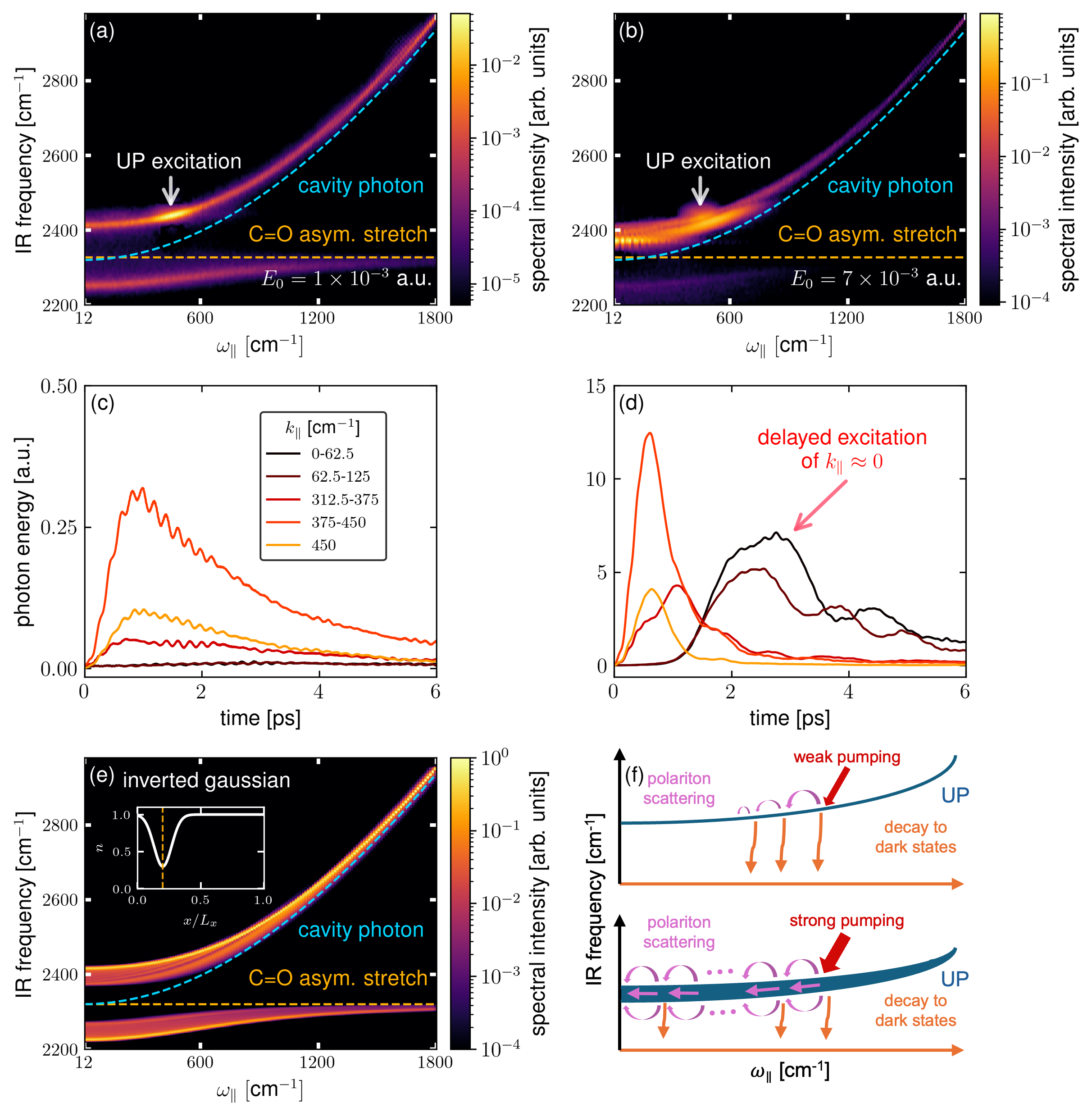}
		\caption{(a,b) Nonequilibrium polariton dispersion relations during  UP transport under weak and strong pumping, corresponding to Figs. \ref{fig:1d_imag}a and c, respectively. (c,d) Associated cavity photon energy dynamics per mode. (e) Broadened polariton dispersion  from an extended Tavis--Cummings model with an  inverse-Gaussian molecular density profile, mimicking the depletion of ground-state molecules at the pulse hot spot. (f) Proposed mechanism for nonlinear freezing of UP transport.
		}
		\label{fig:mechanism}
    \end{figure}

    To identify the origin of UP nonlinear freezing, in Figs. \ref{fig:mechanism}a,b we display the nonequilibrium polariton dispersion relations from the 1D UP pumping simulations at $k_{\parallel}=450$ cm$^{-1}$ (Figs. \ref{fig:1d_imag}a,c). Under weak excitation (Fig. \ref{fig:mechanism}a),  the UPs near $k_{\parallel}=450$ cm$^{-1}$ are strongly excited. The mode-resolved photonic energy dynamics in Fig. \ref{fig:mechanism}c, particularly the delayed excitation of cavity modes at $312.5 \leq k_{\parallel} \leq 375$ cm$^{-1}$, indicates polariton-polariton scattering to neighboring $k_{\parallel}$ values \cite{Li2024CavMD}. However, cavity loss and UP relaxation to dark modes prevent the efficient energy transfer to  the band edge ($k_{\parallel} = 0$); see also a recent relevant work  for exciton–polariton systems \cite{kruppQuantumDynamicsSimulation2025a}.

    Under strong pumping (Fig. \ref{fig:mechanism}b), the  nonequilibrium polariton dispersion relation differs significantly. Here, the UP branch near $k_{\parallel} = 0$ becomes meaningfully excited and broadened with red-shifted frequencies \cite{Dunkelberger2019}. The corresponding mode-resolved photonic energy dynamics in Fig. \ref{fig:mechanism}d also confirm the strong delayed excitation of cavity modes near $k_{\parallel}=0$ (red arrow). Since the polariton group velocity vanishes at small $k_{\parallel}$, this delayed population of the $k_{\parallel}\approx0$ cavity modes is directly responsible for the nonlinear freezing of the UP propagation observed in Figs. \ref{fig:2d_imag} and \ref{fig:1d_imag}.

    We attribute the physical origin of this enhanced $k_{\parallel}\approx 0$ UP population to the pump-induced breaking of in-plane translational symmetry. Under strong localized pumping, the effective oscillator strength for the fundamental $0\rightarrow1$ vibrational transition is reduced near the pulse hot spot due to partial depletion of the vibrational ground-state population \cite{Dunkelberger2019}. In the classical CavMD picture, the same effect manifests as a reduced local linear vibrational susceptibility of anharmonically excited molecules. With the effective oscillator-strength density denoted by $n_{0\rightarrow1}(x, t)$, and noting that the collective Rabi frequency $\Omega_{\rm R}$ scales as $\sqrt{n_{0\rightarrow 1}}$, the pulse pumping generates a spatially localized Rabi contraction $\delta \Omega_{\rm R}(x,t)=\Omega_{\rm R}^{(0)} \{ [n_{0\rightarrow1}(x,t)/n_{0\rightarrow 1}^{(0)}]^{1/2} -1 \}$, which makes the light-matter coupling inhomogeneous along the cavity mirror plane.  
    
    Extended Tavis--Cummings model calculations \cite{Li2024Symmetry} with an inverted-Gaussian inhomogeneous molecular population density (Fig. \ref{fig:mechanism}e) confirms that the resulting in-plane translational symmetry breaking produces significantly broadened polariton branches (as in Fig. \ref{fig:mechanism}b), especially when the magnitude of the Gaussian inhomogeneity becomes large. This symmetry breaking increases the light-matter density of states (i.e., number of eigenstates with meaningful photonic components) at each $k_{\parallel}$ value \cite{Li2024Symmetry}. 
    
    Consequently, as demonstrated in Fig. \ref{fig:mechanism}f, the excited UP can scatter its energy to a greater number of polariton states at progressively smaller $k_{\parallel}$, which significantly enhances the eventual population of the $k_{\parallel}\approx 0$ band-edge modes. By contrast,  in the linear response regime, as only one UP state exists at each $k_{\parallel}$, UP relaxation to dark modes dominates over the cascaded $k_{\parallel}\rightarrow k_{\parallel}'\rightarrow\cdots \rightarrow 0$ polariton-polariton scattering. 
    Because the relaxation from the UP to dark states and photonic leakage due to cavity losses are largely independent of the external pumping strength, their competition with the pump-enhanced polariton-polariton scattering may determine the threshold for this nonlinear freezing pathway.

    A minimal quantum model (SI Sec. III) provides analytical insights of this picture \cite{Li2024Symmetry,Ciuti2003,Carusotto2013}. The pump-induced Rabi contraction acts as a self-induced scattering potential $\delta \hat{H}(t)=\frac{1}{2}\sum_{k,k'}\delta\Omega_{\rm R}(k-k',t)(\hat{a}_k^{\dagger}\hat{B}_{k'} + \hat{a}_k\hat{B}_{k'}^{\dagger})$, where $\hat{a}_{k}^{\dagger}$ ($\hat{B}_{k'}^{\dagger}$) creates a cavity photon (collective molecular bright-state excitation) and $\delta\Omega_{\rm R}(k-k',t)$ is the Fourier transform of the spatial Rabi contraction. Transforming to the polariton basis yields the scattering matrix element $V_{\nu k,\nu' k'}\simeq \frac{1}{2} \delta\Omega_{\rm R}(k-k',t) [C_{\nu k}X_{\nu' k'} + C_{\nu' k'} X_{\nu k}]$ between polariton branches $\nu, \nu' \in \{\text{LP}, \text{UP}\}$ \cite{Carusotto2013,Grochol2008}, where  $C_{\nu k}$ and $X_{\nu' k'}$ are the photonic and vibrational Hopfield coefficients. Because the pump-induced depletion is spatially localized, $\delta\Omega_{\rm R}(k-k')$ carries broad spectral weight, thus enabling large in-plane momentum transfer between polariton states. We note that the CavMD simulations capture polariton-dark-mode coupling and nonperturbative pumping effects, which provide the description of nonlinear polariton dynamics beyond the minimal quantum model.

    Apart from the key role of symmetry breaking,  this nonlinear pathway appears  to also correlate with molecular interactions and anharmonicity. For example, controlled simulations for the harmonic bending mode of \ch{CO_2} forming strong coupling and stretching-mode strong coupling for a dilute gas-phase \ch{CO_2} system (Figs. S6 and S7 of the SI) yield weakened nonlinear freezing, which likely arises from the reduced polariton-polariton scattering for molecular systems resembling closer to independent harmonic oscillators. Regarding the role of cavity lifetimes, the E-field dynamics given in SI Fig. S8 indicate that excessively strong cavity loss (with a lifetime smaller than ~1 ps) may suppress the freezing of the polariton transport.
    
    The nonlinear freezing of vibrational UP transport reported here appears to resemble classical spatial self-trapping \cite{wagnerLargeScaleSelfTrappingOptical1968a} or  polariton self-trapping of exciton-polariton condensates \cite{Abbarchi2013,Ballarini2019,pieczarkaEffectOpticallyInduced2019a,polkingControllingLocalizedSurface2013a}. However, exciton-polariton self-trapping requires condensation and involves a self-induced attractive, polaron-like state stabilized by stimulated scattering \cite{Chestnov2018}, whereas the present mechanism here requires neither condensation formation nor the participation of lower polaritons. Additionally, our finding may relate to the UP condensation reported in exciton–polariton systems\cite{alnatahControlledSwitchingBoseEinstein2026a,chenBoseCondensationUpperBranch2023a}, which requires further investigation.

    From a chemical physics perspective, the UP nonlinear freezing provides a novel mechanism for remote energy transfer between molecules: The spatial location of the molecular energy acceptors can be controlled by tuning $k_{\parallel}$ of the pulse,  directing energy deposition from  nearby to far away from the pulse hot spot (Fig. \ref{fig:remote_energy_transport}d). This photo-induced energy-hopping mechanism differs fundamentally from conventional energy-transfer mechanisms \cite{Forster1948} which predict monotonically diminishing molecular excitations at increased distance from the source.

    \begin{acknowledgement}
This material is based upon work supported by the U.S. National Science Foundation under Grant No. CHE-2502758. This work used the Anvil HPC at Purdue University through allocation CHE250091 from the Advanced Cyberinfrastructure Coordination Ecosystem: Services \& Support (ACCESS) program \cite{Boerner2023}, which is supported by U.S. National Science Foundation grants \#2138259, \#2138286, \#2138307, \#2137603, and \#2138296.

\end{acknowledgement}

\begin{suppinfo}
    Supplementary information PDF file, including additional figures for examining the parameter dependence of UP transport dynamics, simulation details, and the  description of the minimal quantum model.
    2D movies of E-field intensity and \ch{C=O} bond energy distributions at weak and strong pumping.
    The source code and parameters used to generate the simulations available at Ref. \citenum{TELi2026MaxwellLinkGithub}.
\end{suppinfo}
    
 
    \providecommand{\latin}[1]{#1}
\makeatletter
\providecommand{\doi}
  {\begingroup\let\do\@makeother\dospecials
  \catcode`\{=1 \catcode`\}=2 \doi@aux}
\providecommand{\doi@aux}[1]{\endgroup\texttt{#1}}
\makeatother
\providecommand*\mcitethebibliography{\thebibliography}
\csname @ifundefined\endcsname{endmcitethebibliography}  {\let\endmcitethebibliography\endthebibliography}{}

    \clearpage
    

\section*{Supplementary material}

## II. SIMULATION DETAILS

The mesoscale CavMD approach [1] was implemented in the open source software package MaxwellLink [2]. The associated simulation data are available at Github: https://github.com/TaoELi/maxwelllink_examples.

In this work, we performed calculations for simulating both one-dimensional (1D) and two-dimensional (2D) Fabry–Pérot cavities coupled to liquid-phase molecules. In these simulations, the molecules were evenly distributed along a 1D or 2D grid, with each grid point representing a single MD simulation cell consisting of $N_{\mathrm{simu}} = 36$ explicitly simulated $CO_2$ molecules under periodic boundary conditions. For each MD simulation cell, the cell length was set to 25.263 a.u. = 13.368 Å, corresponding to a molecular density of 1.101 g/cm$^{-3}$. The $CO_2$ molecules were modeled by a non-polarizable, anharmonic force field used in pervious CavMD work [3]. In each MD simulation cell, intermolecular Coulombic interactions were calculated using the Ewald summation.

For 1D simulations reported in the main text, the simulation setup consisted of $N_{\mathrm{mode}} = 144$ cavity photon modes coupled to $N_{\mathrm{grid}} = 144$ molecular grid points. The frequency of each cavity mode was determined by

$$\omega_{\mathrm{c}}(l_x) = \sqrt{\omega_\perp^2 + (l_x \Delta\omega_x)^2}, \tag{1}$$

where $\omega_\perp = 2320$ cm$^{-1}$ denoted the fundamental cavity frequency, and the in-plane frequency was $\omega_x = l_x \Delta\omega_x = 12.5\, l_x$ cm$^{-1}$, with $l_x = 1, 2, \cdots, N_{\mathrm{mode}} = 144$ labeling the cavity photon modes and $\Delta\omega_x = c\pi/L_x = 12.5$ cm$^{-1}$ representing the in-plane frequency spacing along the $x$-axis. All the $N_{\mathrm{grid}} = 144$ molecular gird points were located equidistantly along the $x$-oriented cavity plane at positions

$$x_{\mathrm{grid}}^{(i)} = \frac{i \cdot L_x}{N_{\mathrm{grid}} + 1}, \tag{2}$$

for $i = 1, 2, \cdots, N_{\mathrm{grid}}$, given $L_x$ as the cavity plane length along the $x$ direction. As the in-plane frequency spacing was set to $\Delta\omega_x = c\pi/L_x = 12.5$ cm$^{-1}$, the cavity plane length $L_x = 400$ $\mu$m. The effective light-matter coupling strength per molecule was $\widetilde{\varepsilon}_{k\lambda} = 2.5 \times 10^{-5}$ a.u.

For 2D simulations reported in the main text, the simulation setup consisted of $N_{\mathrm{mode}} = 144^2 = 20,736$ cavity modes coupled to $N_{\mathrm{grid}} = 144^2$ molecular grid points. The frequency of each cavity mode was determined by

$$\omega_{\mathrm{c}}(l_x, l_y) = \sqrt{\omega_\perp^2 + (l_x \Delta\omega_x)^2 + (l_y \Delta\omega_y)^2}. \tag{3}$$

Similar as the 1D parameters, here $\omega_\perp = 2320$ cm$^{-1}$; $l_x, l_y = 1, 2, \cdots, 144$; and $\Delta\omega_x = \Delta\omega_y = 12.5$ cm$^{-1}$. The $N_{\mathrm{grid}} = 144^2$ molecular grid points were placed equidistantly on $xy$-plane at positions

$$r_{\mathrm{grid}}^{(i,j)} = \left(\frac{i \cdot L_x}{145}, \frac{j \cdot L_y}{145}\right), \tag{4}$$

where $i, j = 1, 2, \cdots, 144$, and the cavity plane length $L_x = L_y = 400$ $\mu$m. The effective light-matter coupling strength was set to $\widetilde{\varepsilon}_{k\lambda} = 2.083 \times 10^{-6}$ a.u.

The MD simulation cell associated with each grid point was propagated via the modified LAMMPS package [4], which connected to MaxwellLink for self-consistent light-matter dynamics [2]. For equilibrium simulations, each MD simulation cell was initially equilibrated independently for 10 ps under an NVT ensemble at 300 K using a time step $\Delta t = 0.5$ fs. A Langevin thermostat with lifetime 50 fs was applied on the nuclei. After MD pre-equilibration, the coupled light-matter system consisting of many MD simulation cells and cavity modes was propagated under an NVE ensemble using same time step $\Delta t = 0.5$ fs for 20 ps, with a Langevin thermostat of lifetime $\tau_{\mathrm{ph}} = 5$ ps attached solely on photons for representing the cavity loss. For the coupled light-matter dynamics, the initial conditions (positions and momenta) of cavity photon modes were set to obey the Maxwell-Boltzmann distribution at 300 K.

The equilibrium polariton dispersion relation was computed by Fourier transforming the photonic position auto-correlation function using equilibrium NVE trajectories [1]. The nonequilibrium polariton dispersion relation was computed similarly with nonequilibrium pulse-excitation simulations introduced below.

### A. Pulse excitation

For nonequilibrium simulations, in addition to the parameters introduced in the equilibrium simulations, an external pulse was applied to the system. The pulse was defined as

$$\mathbf{E}(\mathbf{r}, t) = E_0\, \mathbf{s}(\mathbf{r}; \mathbf{r}_0, \mathbf{R}) \exp\left[-\frac{(t - t_0)^2}{2\tau^2}\right] \cos[\omega t - \mathbf{k}_\parallel \cdot (\mathbf{r} - \mathbf{r}_0) + \phi]\, \mathbf{e}_\lambda. \tag{5}$$

Here, $\mathbf{r}_0 = (x_0, y_0)$ and $\mathbf{R} = (R_x, R_y)$ denoted the pulse center and spatial extent along $x$- and $y$-axis, respectively; $E_0$ defined the pulse amplitude; and $\mathbf{e}_\lambda$ represented the unit vector along the pulse polarization direction. The Gaussian envelope parameters $t_0 = 0$ and $\tau = 500$ fs specified the temporal center and width of the pulse, respectively. The quantities $\omega$ and $\phi = 0$ denoted the frequency and phase of plane wave part. The spatial phase of the plane wave was determined by the in-plane $k$-vector $\mathbf{k}_\parallel = (k_x, k_y)$.

In the Gaussian pulse defined in Eq. (5), the 2D Hann window $\mathbf{s}(\mathbf{r}; \mathbf{r}_0, \mathbf{R}) = s(x; x_0, R_x) \otimes s(y; y_0, R_y)$ was defined as

$$s(i; i_0, L_i) = \frac{1}{2R_i} \begin{cases} 1 + \cos \frac{2\pi(i-i_0)}{R_i} & , |i| \leq R_i/2 \\ 0 & , |i| > R_i/2 \end{cases}, \quad (i = x, y). \tag{6}$$

In 1D simulations, the center of the pulse located at $x = 0.2\,L_x = 80$ $\mu$m with pulse spatial extent $R_x = 120$ $\mu$m. In 2D simulations, the center of the pulse located at $x = 0.2\,L_x = 80$ $\mu$m and $y = 0.5\,L_y = 200$ $\mu$m with pulse spatial extent $R_x = 0.3\,L_x = 120$ $\mu$m and $R_y = 0.3\,L_y = 120$ $\mu$m.

The pulse defined in Eq. (5) was attached on all cavity photon modes through an external driving term

$$m_{k\lambda}\ddot{\widetilde{q}}_{k\lambda} = -m_{k\lambda}\omega_k^2 \widetilde{q}_{k\lambda} - \widetilde{\varepsilon}_{k\lambda} d_{\mathrm{g},k\lambda} - Q \sum_{\mathbf{r}} \mathbf{E}(\mathbf{r}, t)\mathbf{f}_{k\lambda}(\mathbf{r}), \tag{7}$$

where $\mathbf{f}_{k\lambda}(\mathbf{r})$ and $Q$ denoted the mode function and effective photon charge of the $k$-th cavity photon mode along polarization direction $\lambda$, respectively. Here, the effective photon charge $Q$ was set to $Q = \sqrt{N_{\text{excited}}} Q_{\text{single-mode}}$, where $N_{\text{excited}}$ denoted the number of MD simulation cells at the pulse hot spot, and $Q_{\text{single-mode}} = 0.010$ a.u. represented the photonic partial charge used in single-mode CavMD simulations [5] which corresponded to cavity lifetime $\tau_{\text{ph}} = 5$ ps, yielding $Q$ =0.066 a.u. in 1D and 0.43 a.u. in 2D simulations, respectively. For exciting a specific UP mode, the pulse frequency $\omega$ and in-plane $k$-vector $\mathbf{k}_\parallel$ were selected to match that UP mode. The Gaussian pulse was always turned off at $t_{\text{end}} = 1$ ps.

### B. System size dependence

For 1D simulations, apart from the calculations involving $N_{\text{mode}} = 144$ cavity modes coupled to $N_{\text{grid}} = 144$ molecular grid points, as shown in Fig. S3, we further enlarged the light-matter system size (i.e., $N_{\text{grid}}$ only) by a factor of $2^n$ ($n = 1, 2, 3$) to examine the molecular system size dependence of our findings. During the enlarged simulations, the Rabi splitting was always maintained the same by setting the effective light-matter coupling strength as $\widetilde{\varepsilon} = 2.5 \times 10^5/\sqrt{2^n}$ a.u., and the effective photon charges corresponded to different system sizes were also rescaled as $Q = 0.66\sqrt{2^n}$ a.u. to maintain the same energy gain per molecule.

### C. Absorbing Boundary Conditions

One important feature of our simulations was the use of absorbing boundary conditions to reduce the electric field reflection at the geometry boundary in both 1D and 2D simulations.

The local electric field on the 2D planar Fabry-Pérot cavity at the $\vec{n} = (n_x, n_y)$-th grid point $\mathbf{E}^{(\vec{n})}$ was evaluated as [2]

$$\mathbf{E}^{(\vec{n})} = \sum_{k\lambda} \widetilde{\varepsilon}_{k,\lambda} \left( \widetilde{q}_{k\lambda} + \frac{\widetilde{\varepsilon}_{k,\lambda}}{m_{k\lambda}\omega_k^2} d^{\text{sub}}_{\mathrm{g},k\lambda} \right) \cdot \sqrt{\mathcal{V}} \mathbf{f}_{k\lambda}(\mathbf{r}_{\vec{n}}). \tag{8}$$

In general, for the simulated molecules without permanent dipole moments (e.g. $CO_2$ and $CH_4$), the contribution of the dipole moment term ($d^{\text{sub}}_{\mathrm{g},k\lambda}$) to the local electric field was small. Therefore, as an approximation, the dipole term in Eq. (8) could be neglected, leading to an approximated form of the local electric field:

$$\mathbf{E}^{(\vec{n})} = \sum_{k\lambda} \widetilde{\varepsilon}_{k,\lambda} \widetilde{q}_{k\lambda} \cdot \sqrt{\mathcal{V}} \mathbf{f}_{k\lambda}(\mathbf{r}_{\vec{n}}). \tag{9}$$

Given the 2D rectangular photonic geometry, only the $x$- and $y$-components of $\mathbf{E}^{(\vec{n})} = (E_x^{(\vec{n})}, E_y^{(\vec{n})}, E_z^{(\vec{n})})$ became meaningful. Therefore, we projected $\mathbf{E}^{(\vec{n})}$ on a specific polarization direction (e.g., $y$-axis) as

$$E_y^{(\vec{n})} = \mathbf{E}^{(\vec{n})} \cdot \mathbf{e}_y = \sum_k \widetilde{\varepsilon}_k \widetilde{q}_{ky} \sqrt{\mathcal{V}} \mathrm{f}_{ky}(\mathbf{r}_{\vec{n}}), \tag{10}$$

which could be rewritten in a matrix form:

$$\boldsymbol{E}_y = \boldsymbol{\varepsilon}\boldsymbol{q}_y\boldsymbol{F}_y. \tag{11}$$

Here, $\boldsymbol{E}_y$, $\boldsymbol{q}_y$ and $\boldsymbol{F}_y$ denoted the $y$-component of the local electric field, photonic coordinates, and mode functions, respectively; and $\boldsymbol{\varepsilon}$ was the effective light-matter coupling strength diagonal matrix. The dimensions of the $\boldsymbol{E}_y$, $\boldsymbol{\varepsilon}$, $\boldsymbol{q}_y$, and $\boldsymbol{F}_y$ quantities were $(N_{\text{grid}},)$, $(N_{\text{mode}}, N_{\text{mode}})$, $(N_{\text{mode}},)$ and $(N_{\text{mode}}, N_{\text{grid}})$, respectively.

To introduce the absorbing boundary conditions, we defined the smoothing function $S(x)$ as [6]

$$S(x) = \begin{cases} 0 & , \quad 0 < x < a_0 \\ \left[1 + e^{L(x;a_0,a_1)}\right]^{-1} & , \quad a_0 \leq x \leq a_1 \\ 1 & , \quad a_1 < x < b_0 \\ \left[1 + e^{-L(x;b_0,b_1)}\right]^{-1} & , \quad b_0 \leq x \leq b_1 \\ 0 & , \quad b_1 < x < 1 \end{cases} \tag{12}$$

Here, the dividing fraction $a_0, a_1, b_0, b_1$ satisfied $0 < a_0 < a_1 < b_0 < b_1 < 1$. $L(x; x_l, x_r)$ denoted the linear partition function

$$L(x; x_l, x_r) = \frac{x_l - x_r}{x_l - x} + \frac{x_r - x_l}{x - x_r}, \tag{13}$$

where $x_l, x_r$ were the starting point and ending point of variable $x$, which satisfied $x_l < x < x_r$. For parameters, we used $a_0 = 1/145, a_1 = 0.05, b_0 = 0.95, b_1 = 144/145$.

For applying the absorbing boundary conditions, the smoothing function $S(x)$ was applied along the $x$-axis in the 1D simulations and along both the $x$- and $y$-axes in the 2D simulations. In detail, we applied the smoothing function $S(x)$ on the photonic momenta $\boldsymbol{p}_y$ as an effective "thermostat":

$$\boldsymbol{p}'_y = \boldsymbol{p}_y\boldsymbol{F}_y\boldsymbol{S}\boldsymbol{F}_y^{-1}. \tag{14}$$

In 2D simulations, the smoothing matrix $\boldsymbol{S}(N_x, N_y)$ was evaluated by the product of smoothing function $S(N_x/N_{x,\max})$ along the $x$-direction and $S(N_y/N_{y,\max})$ along the $y$-direction, over all the grid positions, as

$$\boldsymbol{S}(N_x, N_y) = S(N_x/N_{x,\max}) \cdot S(N_y/N_{y,\max}), \tag{15}$$

with the dimension of $(N_{x,\max} = 144, N_{y,\max} = 144)$ satisfying $N_{\text{grid}} = N_{x,\max}N_{y,\max}$, where $(N_x, N_y)$ denoted the molecular gird point index in the 2D rectangular plane. The final smoothing transformation matrix was obtained by flatting the smooth matrix $\boldsymbol{S}(N_x, N_y)$ onto a diagonal matrix $\boldsymbol{S}$ with the dimension of $(N_{\text{grid}}, N_{\text{grid}})$.

## III. MINIMAL QUANTUM MODEL FOR NONLINEAR FREEZING OF POLARITON TRANSPORT

Here, we provide additional details on the minimal quantum model for nonlinear freezing of vibrational polariton transport summarized in the main text. Consider a 1D planar Fabry–Pérot cavity with $M$ discrete photon modes and $M$ molecular grid points, at each grid point $n$, the molecular ensemble is represented by a collective bosonic operator $\hat{b}_n$. Under the rotating-wave approximation, the multimode Tavis–Cummings Hamiltonian reads ($\hbar = 1$):

$$\hat{H} = \sum_k \omega_k \hat{a}_k^\dagger \hat{a}_k + \omega_v \sum_n \hat{b}_n^\dagger \hat{b}_n + \frac{1}{2}\sum_{k,n} \Omega_{\text{R}}^{(0)}\, \varphi_k(x_n) \left(\hat{a}_k^\dagger \hat{b}_n + \text{h.c.}\right), \tag{16}$$

where $\hat{a}_k^\dagger$ creates a cavity photon at frequency $\omega_k$, $\omega_v$ is the molecular vibrational frequency, $\Omega_{\text{R}}^{(0)}$ is the collective vacuum Rabi frequency, and $\varphi_k(x_n)$ is the normalized cavity mode function at grid point $x_n$, satisfying $\sum_n \varphi_k(x_n)\varphi_{k'}(x_n) = \delta_{kk'}$. Under Dirichlet boundary conditions, the mode photonic mode functions are

$$\varphi_k(x_n) = \sqrt{\frac{2}{L}} \sin\left(\frac{\pi k\, x_n}{L}\right), \tag{17}$$

where $L$ is the 1D cavity plane length.

Using the completeness of the mode functions, we define the collective bright-state operator $\hat{B}_k = \sum_n \varphi_k(x_n)\, \hat{b}_n$ for each cavity mode $k$. In the spatially uniform (unperturbed) system, the Hamiltonian becomes diagonal in $k$: each $\hat{a}_k$

couples only to its corresponding $\hat{B}_k$ with strength $\Omega_{\rm R}^{(0)}/2$. Diagonalizing the resulting $2\times 2$ problem at each $k$ yields the UP and LP modes $\hat{p}_{\nu,k} = C_{\nu k}\hat{a}_k + X_{\nu k}\hat{B}_k$ with photonic ($C_{\nu k}$) and vibrational ($X_{\nu k}$) Hopfield coefficients. This provides to the standard description of polariton dispersion relation in planar Fabry–Pérot cavities.

Under strong localized pumping, however, the effective oscillator-strength density $n_{0\to 1}(x,t)$ becomes spatially inhomogeneous due to partial depletion of the vibrational ground-state population at the hot spot. Since $\Omega_{\rm R} \propto \sqrt{n_{0\to 1}}$, this generates a spatially varying Rabi contraction $\delta\Omega_{\rm R}(x,t) = \Omega_{\rm R}^{(0)}\{[n_{0\to 1}(x,t)/n_{0\to 1}^{(0)}]^{1/2} - 1\} \le 0$. Replacing $\Omega_{\rm R}^{(0)} \to \Omega_{\rm R}^{(0)} + \delta\Omega_{\rm R}(x_n,t)$ in Eq. (16) and re-expressing the local operators $\hat{b}_n = \sum_{k'}\varphi_{k'}(x_n)\hat{B}_{k'}$ in the bright-mode basis, the spatially varying Rabi contraction $\delta\Omega_{\rm R}(x,t)$ yields a perturbation of the light-matter Hamiltonian:

$$\delta\hat{H}(t) = \frac{1}{2}\sum_{k,k'}\delta\Omega_{\rm R}(k,k';t)\left(\hat{a}_k^\dagger\hat{B}_{k'} + \text{h.c.}\right), \tag{18}$$

with the mode-overlap matrix element

$$\delta\Omega_{\rm R}(k,k';t) = \sum_n \delta\Omega_{\rm R}(x_n,t)\,\varphi_k(x_n)\,\varphi_{k'}(x_n). \tag{19}$$

If the Rabi contraction were spatially uniform, i.e., when $\delta\Omega_{\rm R}(x_n,t)$ has no spatial dependence, orthonormality would enforce $\delta\Omega_{\rm R}(k,k';t) \propto \delta_{kk'}$. Under this limit, the Rabi contraction only renormalizes the Rabi frequency without introducing intermode scattering. By contrast, given spatial inhomogeneity in $\delta\Omega_{\rm R}(x,t)$, the off-diagonal ($k \neq k'$) couplings between different modes can be generated due to the breaking of the in-plane translational symmetry [7].

Substituting the mode functions in Eq. (17) into (19), and changing from the sum to integral, we find $\delta\Omega_{\rm R}(k,k';t)$ becomes the spatial cosine transform of the spatially dependent Rabi contraction $\delta\Omega_{\rm R}(x,t)$:

$$\delta\Omega_{\rm R}(k,k';t) \approx \frac{1}{L}\int_0^L dx\delta\Omega_{\rm R}(x,t)\cos\left(\frac{\pi(k-k')x}{L}\right) \tag{20a}$$

$$= \delta\Omega_{\rm R}(k-k',t). \tag{20b}$$

Then, transforming Eq. (18) to the polariton basis via $\hat{a}_k = \sum_\nu C_{\nu k}\hat{p}_{\nu,k}$ and $\hat{B}_k = \sum_\nu X_{\nu k}\hat{p}_{\nu,k}$, we further obtain the polariton-basis scattering matrix element

$$V_{\nu k,\,\nu' k'}(t) \simeq \frac{\delta\Omega_{\rm R}(k-k',t)}{2}\left[C_{\nu k}\,X_{\nu' k'} + X_{\nu k}\,C_{\nu' k'}\right], \tag{21}$$

between polariton branches $\nu,\nu' \in \{\text{LP},\text{UP}\}$, as quoted in the main text. Because the pump-induced depletion is spatially localized, $\delta\Omega_{\rm R}(k-k',t)$ carries broad spectral weight ($\Delta k_{\rm max} \sim 1/\sigma$, where $\sigma$ is the spatial width of the depletion) [7]. As such, Eq. (21) provides a perturbative mechanism to allow large in-plane momentum transfer from the initially excited $k_\parallel$ toward the $k_\parallel \approx 0$ band edge where the polariton group velocity vanishes.

---



 \end{document}